\documentclass[fleqn,10pt]{aux/sn-jnl_1785255554_31}
\usepackage[utf8]{inputenc}
\usepackage[T1]{fontenc}
\usepackage{graphicx}	
\usepackage{amsmath}	
\usepackage{amssymb}	
\usepackage{longtable}  
\usepackage{ulem}
\usepackage{newfloat}
\usepackage{setspace}
\usepackage{arydshln}

\usepackage{ragged2e}
\usepackage{lineno}

\DeclareFloatingEnvironment[name={Extended Data Figure},fileext=lof]{extendfigure}
\DeclareFloatingEnvironment[name={Supplementary Figure},fileext=lof]{suppfigure}
\DeclareFloatingEnvironment[name={Supplementary Table},fileext=lof]{suptable}
\DeclareFloatingEnvironment[name={Extended Data Table},fileext=lof]{extendtable}

\newcommand{\actaa}{Acta Astron.}   
\newcommand{\aj}{Astron. J.}   
\newcommand{\apj}{Astrophys. J.}   
\newcommand{\apjl}{Astrophys. J. Lett.}   
\newcommand{\apjs}{Astrophys. J. Suppl. Ser.}   
\newcommand{\aap}{Astron. Astrophys.}   
\newcommand{\memsai}{Mem. Soc. Astron. Italiana}   
\newcommand{\mnras}{Mon. Not. R. Astron. Soc.}   
\newcommand{\nat}{Nature} 
\newcommand{\nar}{New Astron. Rev.}   
\newcommand{\pasa}{Publ. Astron. Soc. Aust.}   

\newcommand{\Tin}{\mbox{$T_{\mathrm{in}}$}}
\newcommand{\Tout}{\mbox{$T_{\mathrm{out}}$}}

\newcommand{\Teff}{\mbox{$T_{\mathrm{eff}}$}}
\newcommand{\logg}{\mbox{$\log g$}}
\newcommand{\Msun}{\mbox{$M_{\odot}$}} 
\newcommand{\Rsun}{\mbox{$R_{\odot}$}}

\newcommand{\gcmcube}{\mbox{$\mathrm{g\,cm^{-3}}$}}
\newcommand{\Ion}[2]{#1{\,\sc#2}}

\begin{document}

\title{A double debris disc around a white dwarf sculpted by two planetary bodies}

\author[1,2]{Nicola Pietro Gentile Fusillo}
\author[3,4]{Felipe Lagos-Vilches}
\author[3,5]{Boris T. G\"ansicke}
\author[6]{Christopher J. Manser}
\author[7]{Detlev Koester}

\affil[1]{Dipartimento di Fisica, Universita` di Trieste, Trieste, Italy}
\affil[2]{INAF-Osservatorio Astronomico di Trieste, Trieste, Italy}
\affil[3]{Department of Physics, University of Warwick, Coventry, UK}
\affil[4]{Departamento de F\'isica, Universidad T\'ecnica Federico Santa Mar\'ia, Av. España 1680, Valpara\'iso, Chile}
\affil[5]{Centre for Exoplanets and Habitability, University of Warwick, Coventry, UK}
\affil[6]{Department of Physics, Astrophysics Group, Imperial College London, London, UK}
\affil[7]{Institut f\"ur Theoretische Physik und Astrophysik, University of Kiel, Kiel, Germany}
\affil[*]{nicolapietro.gentilefusillo@units.it}
\doublespacing

\abstract{Many planetary systems survive the metamorphosis of their hosts stars into white dwarfs,  as demonstrated by the detection of dusty circumstellar discs around these stellar remnants
as well as chemical signatures of planetary debris in atmospheres of these stars. 
A small fraction of these circumstellar discs also contain gas, detected in the form of emission lines from various ionic transitions (most commonly the near-infrared \Ion{Ca}{ii} triplet). All of the 21 known gas discs exhibit double-peaked line profile characteristic of a rotating flat disc.
The mechanism behind the the generation of gas is still debated, but for at least one system it has been linked with the presence of a surviving planetesimal embedded within the dust disc. 
Here we report the discovery of WD\,J132800.85$-$145019.95, a new gas disc system which  exhibits quadruple-peaked instead of double-peaked line profiles. These emission lines originate from two separate gaseous rings extending from 0.14\,\Rsun\ to 0.34\,\Rsun\ and from 0.54\,\Rsun\ to 1.93\,\Rsun. We propose that this structure results from the presence of two distinct planetary bodies within these discs, ranging in size from a few km to the size of Pluto. The detection this stable double-disc presents a new challenge for the current models of 
debris disc formation and evolution.}

\flushbottom
\maketitle
\thispagestyle{empty}

\section*{Main}
\begin{figure}
\centering
\includegraphics[width=0.7\textwidth]{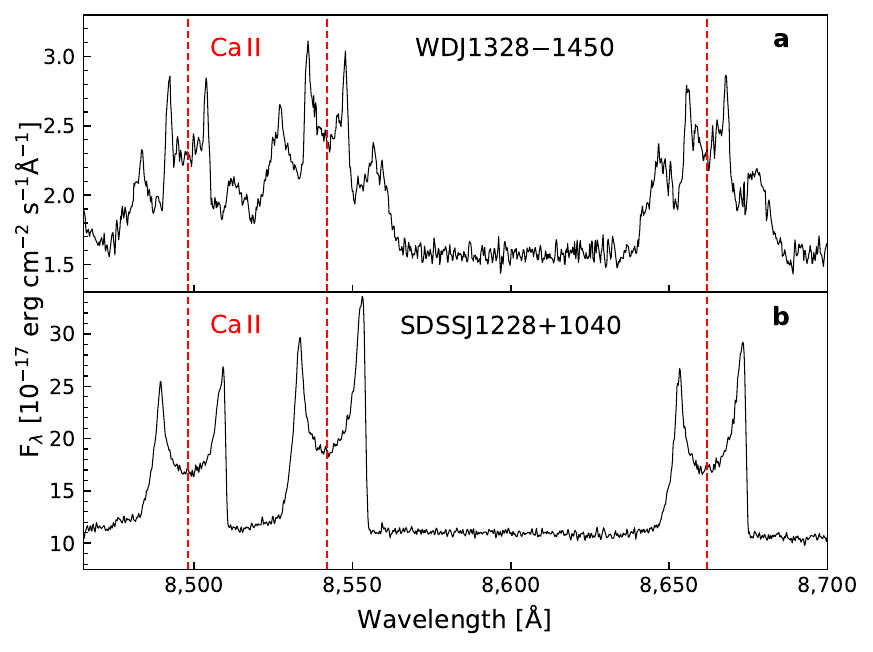}
\caption{\justifying\label{fig:quad_peak}\textbf{Quadruple-peaked emission lines in WD\,J1328$-$1450.} \textbf{a} In contrast to the all 21 gaseous discs published to date, the \Ion{Ca}{ii} 8,498, 8,542, and 8,662\,\AA\  emission lines (shown as red dashed lines) in the X-shooter spectrum of WD\,J1328$-$1450 are quadruple-peaked. \textbf{b} The typical double-peaked emission line profiles of the prototype gas disc system SDSS\,J1228+1040\cite{gaensickeetal06-3d}. The morphology of the emission lines in WD\,J1328$-$1450 indicates the presence of two concentric discs around this white dwarf.}
\end{figure}

We identified an infrared excess in the broad-band photometric  spectral energy distribution (SED) of WD\,J132800.85$-$145019.95 (henceforth WD\,J1328$-$1450) and subsequently obtained medium resolution spectroscopy using the X-shooter spectrograph on the Very Large Telescope (VLT) to probe for atmospheric metal enrichment. The spectra of WD\,J1328$-$1450 contain Ca, Mg and Si absorption lines (Extended Data Table\,1) which unambiguously confirms the accretion of planetary material from a circumstellar dust disc, as well as \Ion{Ca}{ii}, \Ion{O}{i}, \Ion{Mg}{i} and \Ion{Fe}{ii} emission lines from a metallic gaseous disc (Extended Data Table\,2). However, unlike the double-peaked line profiles of all previously known gas discs, the \Ion{Ca}{ii} and \Ion{O}{i} emission lines in WD\,J1328$-$1450 exhibit quadruple-peaked profiles (Fig.\,\ref{fig:quad_peak}).
This unusual morphology of the emission lines can be explained as a superposition of light emitted from two distinct gaseous rings around the white dwarf.

 By comparing the star's spectrum and photometry with appropriate atmospheric models (Extended data Fig.\,1), we determined the star to have an effective temperature of $\Teff=17,424\pm567$\,K, a surface gravity of $\log g=7.76\pm0.19$, and mass of $M=0.47\pm0.09$\,\Msun.

These parameters fall well within the typical range of the 21 other published gaseous disc hosts \cite{gentilefusilloetal21-1d}.
Keeping the stellar parameters fixed, we fitted the metal absorption features in the spectrum of WD\,J1328$-$1450 and measured the photospheric abundances for Ca, Mg and Si (see Methods and Extended Data Table\,1). 

The accretion rates measured for WD\,J1328$-$1450 are unremarkable when compared to the entire known population of debris-enriched white dwarfs \cite{ xuetal19-1, williamsetal24-1}. For comparison, the accretion rates of Mg, Si and Ca in the gas-disc He-atmosphere system SDSS\,J084539.17+225728.0  \cite{wilsonetal15-1} exceed those in WD\,J1328$-$1450 by an order of magnitude. Similarly, the total estimated mass of Ca in the convection zone of WD\,J1328$-$1450 is only 1.53$\times10^{20}$g, a value significantly below what commonly found for similar objects \cite{koester+kepler18}. 
The only unusual property in the photospheric pollution of WD\,J1328$-$1450 is the rather low Mg to Si abundance ratio at 0.2, while in other debris accreting white dwarfs the value is usually between 0.85 and 1.2 \cite{williamsetal24-1}.
This could suggest that the material accreted by the white dwarf could be significantly Mg depleted, but we warrant that our Mg abundance is evaluated from only one line (Extended Data Table\,1) and it is likely blended with an emission feature, leading to possible underestimation. 
\begin{figure}
    \centering
    \includegraphics[width=\hsize]{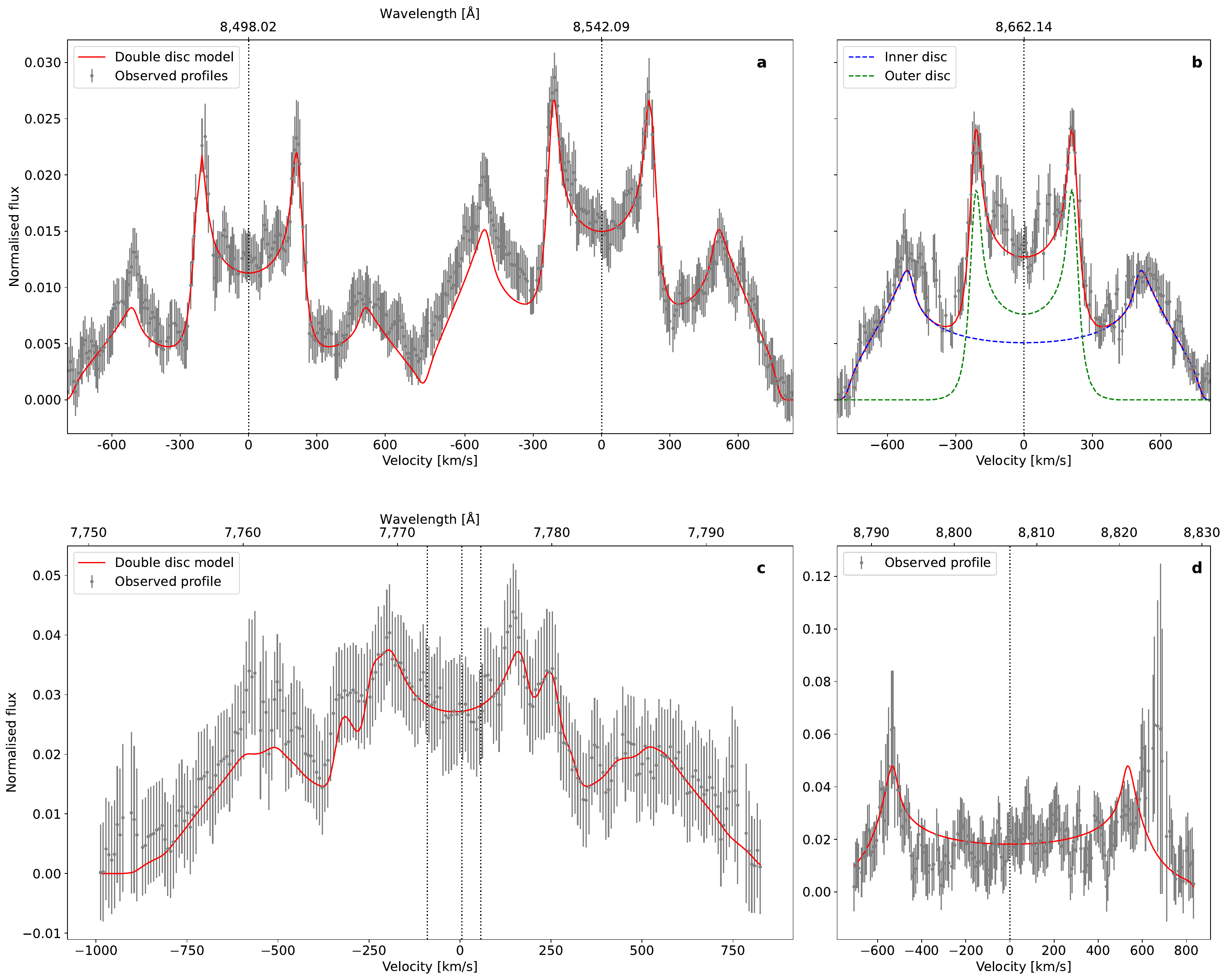}
    \caption{\justifying\textbf{Double-disc model fits to the \Ion{Ca}{ii}, \Ion{O}{i}, and \Ion{Mg}{i} emission lines.} 
    X-Shooter spectra of WD\,J1338$-$1450 (grey points with $1\,\sigma$ error bars) are shown in velocity space relative to the vacuum rest wavelengths (vertical dotted black lines) of the \Ion{Ca}{ii} triplet (\textbf{a, b}), and the \Ion{O}{i} triplet (\textbf{c}) and \Ion{Mg}{i} line (\textbf{d}). The red lines show the best-fit double-disc models (see Extended Data  Table\,3 for more details). \textbf{b}, The individual contributions of the inner (blue) and outer (green) discs are also shown for the \Ion{Ca}{ii} 8,662\,\AA\ component.  The inner disc extends to larger velocities, reflecting its higher Keplerian motion. A small asymmetry in the peak heights suggests a slightly eccentric orbit. \textbf{c}, The morphology of the \Ion{O}{i} 7,774\,\AA\ triplet resembles that of the \Ion{Ca}{ii} triplet although the blending of its  three components results in a more complex structure. Again, some asymmetry is observed indicating a slightly eccentric inner disc. \textbf{d}, The \Ion{Mg}{i} 8807\,\AA\ line is only detected in emission from the inner disc. The red wing of this line is affected by large uncertainties in the data}
    \label{fig:Ca2_triplet_best}
    \end{figure}
Though our abundance measurements are subject to some uncertainties they indicate that, WD\,J1328$-$1450, despite its bright and dynamically active debris disc, does not accrete from it at a high rate. Alternatively, it is possible that the debris disc surrounding WD\,J1328$-$1450 only formed relatively recently and the white dwarfs has only been accreting for one or two diffusion timescale (for the past $3\times10^{5}$ - 6$\times10^{5}$ yr). In this scenario accretion onto the white dwarfs has yet to reach a steady state rate, leading to underestimated mean accretion rates. This would support the idea that we are witnessing a recent disruption event and that the peculiar disc structure of WD\,J1328$-$1450 may result from a chaotic early stage in the disc formation.

\begin{figure}
\includegraphics[width=0.65\hsize]{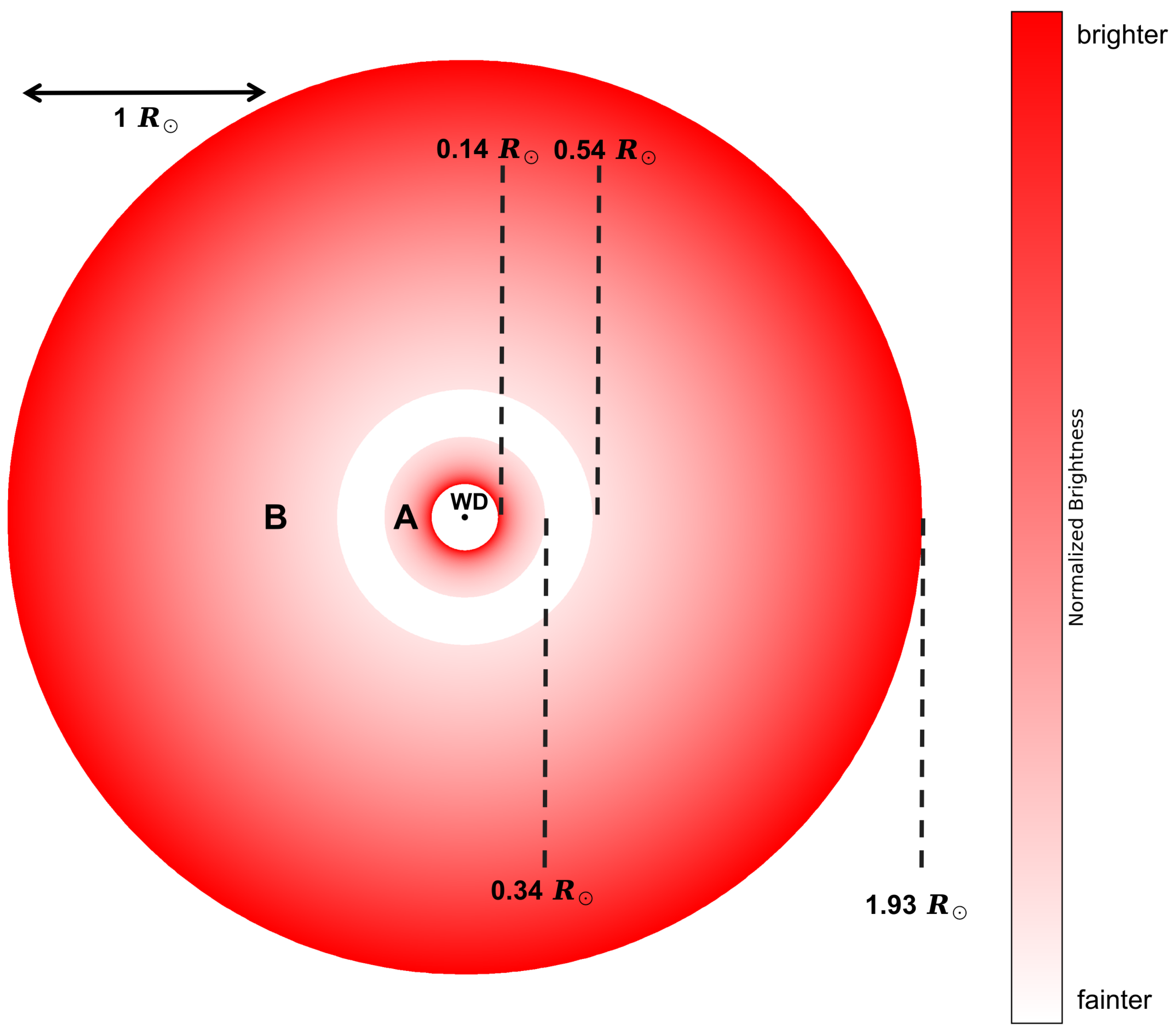}
\centering
\caption{\justifying\label{fig:dik_cartoon} \textbf{Schematic top-down view of the radial location of gas based on the line emission profile modelling.} The coloured regions indicate the location of the observed Ca II triplet emission. The double disc structure around WD\,J1328$-$1450 (inner disc A, outer disc B) and the gap are shown to scale. A 1\,\Rsun bar in the top left corner is shown for size reference. The colour gradient reflects the brightness profile of the discs.}
\end{figure}

We modelled the emission line profiles of \Ion{O}{i}, \Ion{Mg}{i} and \Ion{Ca}{ii} as originating from two concentric and co-planar circular gaseous discs assuming that their surface brightness, as a function of the distance from the white dwarf, follows a power law (Fig.\,\ref{fig:Ca2_triplet_best}). The nearly symmetric profile of the emission lines has remained unchanged for more than 2 years supporting the hypothesis that the gas discs are close to circular and uniform. Under these assumptions, the shape of the \Ion{Ca}{ii} triplet is best reproduced with two discs extending from $0.14^{+0.01}_{-0.02}$\,\Rsun\ to $0.34^{+0.02}_{-0.05}$\,\Rsun\ and from $0.54^{+0.44}_{-0.16}$\,\Rsun\ to $1.93^{+0.10}_{-0.30}$\,\Rsun, with an inclination with respect to the line of sight $i=77^{+9}_{-13}$\,degrees. Interestingly, the surface brightness profile of both discs is quite different:  while the emission from the inner disc decreases with distance, the opposite occurs with the outer disc (Fig.\,\ref{fig:dik_cartoon}, Extended data Table\,3). Our model slightly underestimates the intensity of the blue wing of the nested profiles as shown in Fig. \ref{fig:Ca2_triplet_best}. This discrepancy may indicate that the inner disc either deviates from circular symmetry with moderate ellipticity, or exhibits a non-uniform gas distribution.
However, since the peaks remain symmetric in velocity around the line center in all the observations, we do not find evidence for disc precession which would imply an eccentric orbit. Therefore, the observed flux asymmetry most likely suggests a non-uniform azimuthal distribution of gas in the inner disc, possibly associated with clumpy structures.
Because of the lower signal-to-noise-ratio and blending of the individual line components, modelling of the \Ion{O}{i} emission triplet is less constraining than the results obtained from the \Ion{Ca}{ii} lines, but it nonetheless results in a consistent double disc configuration (see Methods).  In contrast, the \Ion{Mg}{i} line shows a double peak profile indicative of emission originating only from the inner gas disc (Fig.\,\ref{fig:Ca2_triplet_best}\,\textbf{d}).

\begin{figure}
\centering
\includegraphics[width=\hsize]{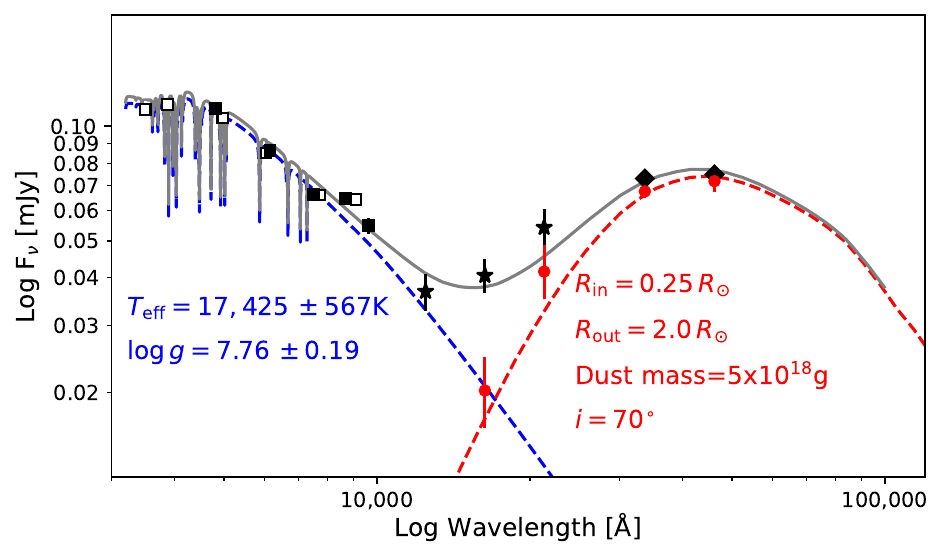}
\caption{\justifying\label{fig:SED} \textbf{The infrared excess of WD\,J1328$-$1450.} 
The best-fit white dwarf model for WD\,J1328$-$1450 (dashed blue line, Extended Data Table  \,3) is compared to the photometry from Pan-STARRS, Skymapper, VHS, and \textit{WISE} (black full squares, empty squares, stars and diamonds respectively, with $1\,\sigma$ errorbars).The photometric IR excess calculated as IR photometry minus the contribution from the white dwarf in the same bands, is represented by the red points. We modelled this IR excess using MCFOST as a disc having scale height at the inner edge equal to the white dwarf radius ($0.015\,\Rsun$), moderate flaring, and outward-increasing dust density (dashed red line.}  
\end{figure}

If the inner and outer discs shared the same Ca/Mg ratio, we would detect Mg emission from the outer ring at $>6\sigma$ significance (Extended Data Fig.\,2). The absence of this detection provides direct, model-independent evidence that Mg is significantly depleted in the outer ring relative to the inner ring.
This compositional difference is an important clue given that different planetary materials exhibit distinct Mg/Ca ratios: chondrites range from Mg/Ca $\simeq 5-15$, Earth's mantle has Mg/Ca$ \simeq 8$, while Earth's crust shows Mg/Ca $\simeq1.6$ \cite{McDonough2003,RudnickGao2003,Lodders2003}. The marked difference between the two rings strongly suggests they originate from distinct parent bodies with different formation histories, or alternatively from different structural layers of a differentiated planetesimal (e.g., Mg-rich mantle material versus Mg-depleted crustal material).
We note that, whereas the model of two discs separated by a gap results in a good fit to the observed quadruple-peaked emission lines of WD\,J1328$-$1450, a single disc in which a vertical extension results in self-shadowing of the region $R\simeq0.34-0.54$\,\Rsun\ would lead to a similar outcome. However, that scenario is not physically backed up by any existing theoretical model of the debris discs around white dwarfs (see Supplementary Information).
The dusty component of the debris disc is detected in the form of a strong infrared excess above the flux of the white dwarf (Fig. \ref{fig:SED}). The fractional infrared flux excess of the dust disc around WD\,J1328$-$1450, i.e. the ratio of the flux emitted by the disc and the stellar flux, is $\tau = 2.7$\,\%, making this one of the brightest dusty debris disc published to date \cite{gentilefusilloetal21-1d,dennihyetal20-2d, rocchettoetal15}. 
We modelled the infrared excess using the radiative transfer code MCFOST \cite{pinteetal06, pinteetal09} and obtained a satisfactory fit with a disc having an inner radius of $0.25\,\Rsun$, outer radius of $2\,\Rsun$, scale height at the inner edge equal to the white dwarf radius ($0.015\,\Rsun$), moderate flaring, and outward-increasing dust density. Although the model reproduces the observations, we note that several fit parameters are degenerate and alternative disc configurations may fit the data equally well. Nonetheless, our analysis supports the growing observational \cite{swanetal20-1} and theoretical evidence that dusty debris discs around white dwarfs can develop substantial vertical extensions \cite{malamudetal21-1,Kenyon+Broomley2017} due to dynamical excitation \cite{Kenyon+Broomley2017}. The dust discs are expected to be optically thin, as demonstrated by prominent silicate emission features \cite{reachetal09-1, reachetal25-1} and observations of partial transits \cite{vanderburgetal15-1}.

Thermal dust emission cannot provide detailed information on the radial structure of the disc due to its broad blackbody spectral kernel. In contrast the sharp emission lines of gaseous discs contain rich dynamical information on the geometry of the emitting regions. Therefore the quadruple-peaked emission profiles provide strong evidence for two distinct gas discs, whereas the dust continuum flux offers no comparable constraint on the spatial distribution of solid material.
One might hypothesize a scenario where a single planetary body opens a gap in both gas and dust, creating two separate discs. However, this would reduce the emitting surface area, making the system's large IR excess even harder to explain. More fundamentally, a solid body within a dust-depleted gap could neither clear the gas from that region nor provide a mechanism for generating the gas in the two discs (see Supplementary Information).

The two concentric gas discs detected at WD\,J1328$-$1450 require both a mechanism that generates the gas within each of them, and avoids the radial spreading of the gas which would close the gap between them. Evidence of strongly localised gas has been detected in the gas disc of WD\,J0846+5703, where the sharp double-peaked emission corresponds to a narrow ring only extending from $\simeq1.0$ to $\simeq1.1$\,\Rsun, for an adopted inclination of $21$ degrees \cite{gentilefusilloetal21-1d}. The most likely mechanism limiting the radial spreading of gas is the sublimation-condensation equilibrium. After gas is produced either by dust collision or sublimation, it spreads viscously on timescales $\tau_\mathrm{visc} \sim$ (10-20)/$\alpha$ years \cite{Metzer_2012-1}. However, as the gas moves outward and cools, it re-condenses below T $\simeq 1,200-1,500$ K \cite{swanetal19-1, rafikov_2012-1}, creating a confined gas ring near the sublimation source. The re-condensation timescale is significantly shorter than the viscous spreading timescale \cite{Metzer_2012-1}, limiting the size of the gas disc and in the case of  WD\,J1328$-$1450, preventing significant radial mixing between the two rings .

Dust sublimation at the inner edge of the disc \cite{Metzer_2012-1,Rafikov_2011-1}, and a collisional cascade of dust particles \cite{Kenyon_2017-1,Kenyon_2017-2,Swan_2021-1} are both alternative hypotheses put forward to explain the origin of gas in white dwarf debris discs. However the double gas disc configuration of WD\,J1328$-$1450 represents a challenge for both these interpretations. The idea that all the gas could be generated by sublimation at the inner edge of the dust disc is incompatible with the existence of a gap in the gas radial distribution. Similarly, in the collisional cascade scenario, the existence of the gap would require some unknown mechanism capable of inhibiting collisions and gas spreading, but only in a limited area of the disc. 
Another possibility is that WD\,J1328$-$1450 is an unresolved double white dwarf system with each component hosting its own gaseous debris disc. However, this scenario would require some extremely fine-tuned initial conditions for the existence of the planetary systems (see Supplementary Information) and, more importantly, the chances of both white dwarfs independently being in a state that allows a gaseous debris disc to persist are extremely small. \cite{manseretal20-1}
Therefore, another scenario for gas-generation in two discs needs to be invoked.

In the archetypal gas-disc white dwarf systems SDSS\,J1228+1040 \cite{gaensickeetal06-3d} the double-peaked emission lines arise from a relatively narrow and eccentric gaseous disc that precesses with a period of $\simeq 25$\,years. This precession period is 100,000s times longer than the orbital period of the gas, and the fact that the radial and azimuthal intensity distribution of the gas emission remains constant over decades requires a mechanism that produces the gas in a stable geometry. Furthermore the detection of a periodic  $\simeq2$\,h signal in the emission lines shape and strength in SDSSJ1228+1040 can only be explained by the presence of a dense solid body embedded in the disc generating gas by interaction with the dust \cite{manseretal19-1}.

Since the double disc structure about WD\,J1328$-$1450 requires stable gas generation at two distinct and confined radial locations, we argue that dust interaction with a planetesimal analogous to SDSS\,J1228+1040, is the most likely origin of the gas, and two gas discs would therefore require two planetesimals on concentric orbits.

Though we cannot determine the exact orbits of the two theorized planetesimals orbiting WD\,J1328$-$1450, the inner and outer edges of the two gas discs provide some stringent limits: the inner planetesimal must be located between 0.14\,\Rsun\ and 0.34\,\Rsun\ and outer one between 0.54\,\Rsun\ and 1.93\,\Rsun. In order to survive tidal disruption at such close distances from the white dwarf, a relatively large body must have an extremely high density or sufficiently high internal strength \cite{manseretal19-1}. Assuming that the body embedded in the inner disc analogous has no internal strength, it would need to be have a density higher than any known material,  ranging from $\gtrsim 3,600\,\gcmcube$ at the inner edge to $\gtrsim 252\,\gcmcube$ at the outer edge. In the outer disc the highest required density is $\gtrsim 63\gcmcube$, over three times that of the densest exo-planet \cite{toledoetal20-1}, and physically plausible  values are only found closer to the outer edge. Therefore, we conclude that any planetesimals surviving within the discs must possess some degree of internal strength.

Using typical internal strengths for rocky Solar System bodies, we can estimate the largest planetesimal that could avoid tidal disruption at the distances of the discs. For an iron-like density ($\rho_\mathrm{Fe} = 7.8,\gcmcube$), a planetesimal in the inner disc could a have maximum diameter of 45 km near the inner edge and 172 km near the outer edge. In the outer disc, the corresponding limits increase to about 346 km at the inner edge and 2,340 km at the outer edge. Because the density requirement in the outer disc is more relaxed, a rocky object with moderate internal strength ($\rho \approx 3.5,\gcmcube$) could also remain intact there, allowing maximum sizes up to roughly 3,500 km at the outer boundary.

To put these numbers into perspective, while the upper size of the inner planetesimal would be that of a typical asteroid in the asteroid belt of our Solar System, the maximum size of the outer planetesimal could range from that of Ceres (the largest asteroid in our Solar System) to that of Pluto or even the Moon, depending on the distance from the white dwarf. 

Although gaseous debris discs have been known for two decades, our understanding of their formation, architecture and long-term evolution remain poorly understood. The discovery of a stable double debris disc around WD\,J1328$-$1450 provides additional observational constraints for the development of physically more detailed models for these discs. Our analysis suggests that two dense planetary bodies are required to sustain this double disc, strengthening the idea that gaseous debris discs may generally arise from interactions between dust and planetesimals that, at least for some time, survive the tidal disruption process. The ring-like morphology of this system is strikingly similar to structures commonly seen in protoplanetary discs\cite{vandermareletal18}. Despite clear differences in scale and in the nature of the central star, WD\,J1328$-$1450 provides evidence that similar processes may shape both the formation and the final evolution of planetary systems.

\clearpage

\section*{Methods}

\subsection*{X-Shooter observations and data reduction}
WD\,J1328$-$1450 was identified as a white dwarf candidate based on its Gaia DR2 parallax and colour \cite{gentilefusilloetal19-1} and flagged as a potential debris disc host based on the infrared excess detected in its VHS and WISE photometry. We observed the star using X-Shooter, an intermediate resolution echelle spectrograph mounted on the Very Large Telescope at Paranal. On 2021 May 15 we obtained three spectra with exposure times of 1,800\,s and 1,750\,s in the UVB and VIS arms, respectively. We observed the star again on 2021 June 28 and 2021 July 2 obtaining five spectra with exposure times of 1,171\,s and 1,200\,s in the UVB and VIS arms, respectively. And finally, we obtained 2 more spectra on 2023 May 12 with exposure times of 1,270\,s and 1,300\,s in the UVB and VIS arms, respectively. All observations were taken using 1.0 and 0.9\,arcsec slit widths in the UVB and VIS arms, respectively. All data were reduced using the standard procedures within the REFLEX (version 2.9.1, http://www.eso.org/sci/software/esoreflex) reduction tool developed by the European Southern Observatory (ESO). Telluric line removal was performed on the reduced spectra using MOLECFIT (version 1.1.0). In addition to the He lines, the X-Shooter spectra shows absorption lines from \Ion{Ca}{ii}, \Ion{Mg}{ii}, and \Ion{Si}{ii}; and emission lines of \Ion{Ca}{ii}, \Ion{O}{i},\Ion{Mg}{ii} and \Ion{Fe}{ii}. 

In addition to spectroscopy, for the analysis of WD\,J1328$-$1450 we used a range of data from wide-area surveys: the parallax measurement from \textit{Gaia} Early Data Release 3 (DR3) \cite{gaia:2021} and photometry from SkyMapper Data Release 2 (DR2) \cite{skymapperDR2},  Pan-STARRS DR2 \cite{panstarrs_survey}, Vista Hemisphere Survey DR6 (VHS) \cite{VHS13-1}, and catWISE2020 \cite{catWISE} (see Supplementary Table\,1). We propagated the \textit{Gaia} coordinates of WD\,J1328$-$1450 to the different epochs of the survey observations using its \textit{Gaia} EDR3 proper motions, and then used a matching radius of three arcsec to retrieve the astrometry and photometry. We did not include the \textit{Gaia} EDR3 photometry in our analysis because of its extremely broad band-passes. 

\subsection*{White Dwarf Parameters}
\label{sec:atm}
We modelled the spectroscopy and photometry of WD\,J1328$-$1450 using white dwarf atmosphere models \cite{koester:2010} to determine the main atmospheric parameters, i.e. the effective temperature (\Teff) and the surface gravity (\logg). Standard techniques to measure these parameters involve fitting either the normalised He line profiles \cite{bergeronetal11-1} or the photometry and the parallax \cite{gentilefusilloetal21-2c}, and both methods proved challenging in the case of WD\,J1328$-$1450. 

The many broad and blended emission lines result in distortions of the continuum level flanking the He absorption lines, significantly affecting the normalisation process. Given that $\log g$ is particularly sensitive to such distortions of the continuum level resulted in a poorly constrained range $\log g=7.3$ to 8.7. A fit to the photometric data and parallax of the star is prone to uncertainties due to contamination by the infrared flux from the dusty debris, a large parallax errors and a correspondingly uncertain reddening, resulting in atmospheric parameters which diverge from those obtained from the spectroscopic fit. 

We therefore adopted an iterative fit, where we determine \Teff\ from the spectroscopic data and \logg\ from the photometric and parallax data. This approach converges on a solution at $\Teff=17,424\pm567$\,K, and $\log g=7.76\pm0.19$ independently of the initial values. Adopting the mass-radius relation for a thin H envelope \cite{bedard20-1} we estimate the mass of white dwarf at $M=0.47\pm0.09$\,\Msun\ and a cooling age of $\tau_\mathrm{cool}=73_{-26}^{+49}$\,Myr.

\subsection*{Photospheric Metal Abundances}
\label{sec:abundances}
Fixing the stellar parameters obtain from this fit, we then computed synthetic spectra for WD\,J1328$-$1450 which also include metals. Although the X-Shooter spectrum only reveals absorption lines of Si, Mg, and Ca, we included C, O, Na, Al, P, S, Ti, and Cr with abundances scaled relative to Ca and adoption a Bulk Earth composition. The measured abundances of Mg, Si, and Ca are provided in Extended Data Table\,1. Several of the metal lines are blended with emission features (Extended Data Table\,2) which considerably complicated the fitting procedure (Extended Data Fig.\,1), and hence our model does not perfectly replicate all absorption line profiles. The complexity of the underlying emission lines introduces systematic uncertainties that are difficult to quantify, and we stress that the uncertainties quoted in Extended Data Table\,1 should be taken as estimates. Taking the photospheric abundances at face value, we computed the diffusion time scales of the elements included in our analysis, their downward fluxes assuming accretion-diffusion equilibrium, and the masses of metals contained within the convection zone (Extended Data Table\,1).

\subsection*{Emission lines and double disc model.}
\label{sec:e_lines}
We model the nested double-peaked shape of the emission lines (Supplementary Table\,2) by assuming two concentric and coplanar circular gaseous discs with surface brightness profiles given by a power-law of the form $f(r) \propto r^{-\alpha}$, $r$ being the distance from the white dwarf centre \cite{Smak_1981}. If suffixes $n=A, B$ denote the inner and outer discs respectively, then the quadruple-peaked emission line profile, $f_\mathrm{quad}(v_\mathrm{obs})$,  as function of the observed Doppler velocity broadening $v_\mathrm{obs}$, is given by

\begin{equation}
    f_\mathrm{quad}(v_\mathrm{obs})= \int_{R_{\mathrm{in},A}}^{R_{\mathrm{max},A}} \frac{r^{3/2-\alpha_A} \,dr}{V^2_\mathrm{K}(r)\sin^{2}(i)-v_\mathrm{obs}^2} +   \int_{R_{\mathrm{in},B}}^{R_{\mathrm{max},B}} \frac{r^{3/2-\alpha_B} \,dr}{V^2_\mathrm{K}(r)\sin^{2}(i)-v_\mathrm{obs}^2},
\end{equation}
with $V_\mathrm{K}(r)$ being the Keplerian velocity at a distance $r$ from the white dwarf, $R_{\mathrm{in},n}$ the locations of the inner edges of the discs,  
\begin{equation}
    R_{\mathrm{max},n}=\min\left( \frac{GM_\mathrm{WD}\sin^2(i)}{v^2_\mathrm{obs}} R_{\mathrm{out, n}}\right)
\end{equation}
the upper integration limit, $G$ the gravitational constant, $M_\mathrm{WD}$ the mass of the white dwarf, $R_{\mathrm{out},n}$ the outer edge of the discs and $i$ the inclination of the disc, where we define $0^{\circ}$ as a face-on disc and  $90^{\circ}$ as an edge-on disc. The resulting profile in velocity space is then convolved with a Gaussian of full width at half maximum $FWHM_v=c/R$, where $R=8,900$ is the spectral resolution of the X-Shooter VIS arm and $c$ the speed of light, to simulate the instrumental response. The profile is then normalised by its total flux $F_\mathrm{nested}=F_A+F_B$, with $F_{n=A,B}$ the flux contributions from the inner and outer discs, respectively. We fix $M_\mathrm{WD}=0.47 M_{\odot}$ (see Methods, White Dwarf Parameters). The observed nested profiles are fully described by nine parameters: $R_{\mathrm{in},A}$, $R_{\mathrm{out},A}$, $R_{\mathrm{in},B}$, $R_{\mathrm{out},B}$, $\alpha_A$, $\alpha_B$, $i$, the relative contributions of the two discs $\mathrm{\zeta}=F_B/F_A$, and the systemic velocity $v_\mathrm{sys}$ resulting in an overall Dopper-shift of the observations with respect to the laboratory wavelengths of the \Ion{Ca}{ii} triplet. 

Using a Bayesian approach, we first fit our model to the calcium triplet by maximizing the logarithm of the probability of the data given the model (i.e. the likelihood). Since each emission line in the calcium triplet is expected to be produce under the same physical conditions, we use a single value for $\mathrm{R_{in,A}}$, $\mathrm{R_{out,A}}$,$R_\mathrm{in,B}$,$\mathrm{R_{out,B}}$, $\alpha_A$ and $\alpha_B$. To account for the partial blending of the \Ion{Ca}{ii} lines at 8,498 and 8,542\,\AA\ we included the parameter $K_\mathrm{blend}$ defined as the flux ratio between both lines. This leaves a total of 12 free parameters listed in Extended Data Table\,3. The parameter space was sampled with the affine-invariant Markov chain Monte Carlo (MCMC) algorithm implemented in the python package \texttt{emcee} using 90 walkers with a chain length and burn-in of $50\tau_\mathrm{max}$ and $5\tau_\mathrm{max}$ iterations respectively, with $\tau_\mathrm{max}$ being the maximum autocorrelation time between the 12 parameters. To determine the value of $\tau_\mathrm{max}$, the autocorrelation time was calculated every 50 steps until its estimate changed by less than 1\,\%.

From the derived posterior distributions shown in Supplementary Fig.\,1, the radial extension of the inner gaseous disc goes from $\approx0.14$ to $\approx0.34$\,\Rsun, while the outer disc extends from $\approx0.54$ to $\approx1.93$\,\Rsun. Our results also suggest a value above 30 degrees for the inclination of the discs, with a mean value of $77$ degrees. 

Unlike the calcium triplet, the blending of the oxygen triplet at $7,774$\,\AA\, did not allow to obtain a good characterization of the inner and outer discs as the fitting becomes highly degenerate (Supplementary Fig.\,2). For this reason, we fixed the inclination and systemic velocity to the values obtained from the fit of the calcium triplet. As shown Fig. \ref{fig:Ca2_triplet_best}, the double-disc model is able to reproduce the main features of the observed profile, and the extension and brightness profile derived for the inner disc are in agreement to those found for the calcium triplet (see Extended Data Table\,3). Conversely, the outer disc parameters are less constrained as their estimates are derived from the central region of the triplet, where the double-peaked structures are more difficult to disentangle. 

\noindent
In addition to both triplets, we attempted to fit the \Ion{Mg}{i} line at 8,806\,\AA\ using a single-disc model, setting the inclination and systemic velocity as was done for the oxygen triplet (Supplementary Fig.\,3). The derived inner disc extension and brightness profile are in good agreement with the above estimates considering its low signal and the possible contamination from a \Ion{Mg}{ii} line at 8,824.318\,\AA, i.e. next to the red-shifted peak of the \Ion{Mg}{i} line.

\subsection*{Lack of circumstellar absorption}
At the inferred inclination of $i=77^{+9}_{-13}$\,degrees, one might expect to detect circumstellar absorption if the gas disc were geometrically thick enough. However, the scale height of a gas disc in thermal equilibrium with the white dwarf radiation is $H/r\simeq c_{s}/v_{K}$. Where $H$ is the height of the disc and $r$ is its radial extension, $c_\mathrm{s}$ is the sound speed and $v_\mathrm{K}$ is the Keplerian velocity. Assuming gas temperatures of $\simeq 3,000$\,K and $\simeq 1,300$\,K for the inner and outer discs respectively (based on radiative equilibrium), and Ca-dominated composition (molecular mean mass $\mu \simeq 40$), we estimate $H/r = 0.005-0.008$. For circumstellar absorption to be detectable, the line of sight must intersect the disc material, requiring  $ i> (90$ \,degrees $-\arctan(H/r)) = 89.6$\,degrees. The absence of circumstellar absorption therefore only constrains the inclination to $i< 89.6$\,degrees, consistent with our emission line modelling ($i=77^{+9}_{-13}$\,degrees). This also implies that the disc cannot have an optically thick component geometrically thicker than predicted by thermal equilibrium, with H/r $\lesssim 0.49$ (corresponding to our lower inclination limit of 64\,degrees).

\subsection*{Evidence of differentiated composition}
The excitation energy for the \Ion{O}{I} 8446\,\AA\ line  is 10.74\,eV, while for \Ion{Mg}{I} at 8806\,\AA\ is only 5.93 eV—approximately half the energy. Since we clearly detect \Ion{O}{I} emission from both the inner and outer discs, the excitation conditions in the outer disc must be more than sufficient to produce \Ion{Mg}{I} emission if magnesium were present at comparable abundances. 
To quantify the expected \Ion{Mg}{I} emission from the outer disc, we use the \Ion{Ca}{II} triplet as a reference. The four-peaked \Ion{Ca}{II} profiles reveal that the outer disc produces, on average, 1.22 times more flux than the inner disc. If we assume the outer and inner discs share the same Ca/Mg abundance ratio, we can predict the expected \Ion{Mg}{i} peak flux from the outer disc by scaling the observed inner disc \Ion{Mg}{I} peak flux (1.94 $\times 10^{-17}$ erg s$^{-1}$ cm$^{-2}$ \AA$^{-1}$) by this factor, yielding an expected peak flux of 2.37$\times 10^{-17}$ erg s$^{-1}$ cm$^{-2}$ \AA$^{-1}$. 
However, at the wavelengths where we would expect \Ion{Mg}{i} emission from the outer disc, we measure only continuum emission from the inner disc with an average flux of (1.58 $\pm$ 0.13)$\times 10^{-17}$ erg s$^{-1}$ cm$^{-2}$ \AA$^{-1}$. The predicted \Ion{Mg}{i} peaks would stand $\simeq6\sigma$ above this continuum level if magnesium were present at inner-disc abundances (Extended Data Fig.\,2). The complete absence of detectable \Ion{Mg}{i} emission from the outer disc therefore provides, model-independent evidence that magnesium is significantly depleted in the outer disc relative to the inner disc.

\subsection*{Infrared excess and circumstellar dust}
\label{sec:IRE}
Using the best fitting model for the white dwarf we can predict the infrared flux of the isolated white dwarf and use this to characterize the infrared excess due to the thermal emission of the circumstellar dust. We begin by treating this system in a similar way to all other know gas-disc hosts, assuming that the infrared excess can be described as originating from a single dusty debris disc. We adopt the model of a passive, geometrically thin, optically-thick disc model from \cite{jura03} (see e.g.\cite{rocchettoetal15,vonhippeletal07, kilicetal12}). Free parameters in this model are the: temperatures at the inner and outer edges of the disc, \Tin\ and \Tout, and disc inclination, $i$. Notably, a significant degeneracy exists \cite{girvenetal12-1} between \Tin\ and $i$  and without photometric measurements red-wards of 4.6\,$\mu\mathrm{m}$,  \Tout\, remains poorly constrained. The best fit we achieve with this model is for a disc  with $\Tin=1,943\pm102$\,K, $\Tout=694\pm153$\,K and $i=0\pm4$\,degrees. These temperature corresponds approximately to a minimum inner edge of 0.10\,\Rsun\ and a maximum outer edge of 0.65\,\Rsun. The outer edge does not represent the physical limit for the presence of dust as material beyond this limit can exist, but it would be too cool to significantly contribute to the observed infrared flux. Despite the fact that this model of a geometrically thin, optically-thick disc is able to correctly reproduce the infrared excess, it does not necessarily provide a realistic physical interpretation.  In fact with $i=0$\,degrees the dust disc would be face-on to the observer and, assuming the dust is co-planar with the gas, this configuration would not produce observable multi-peaked emission lines. For comparison, we recalculated the flat dust disc model this time fixing the parameters to those obtained by modelling the emission lines from the gas-discs, i.e. lower inclination limit of $i=64$\,degrees, inner edge 0.14\,\Rsun\ and outer edge 1.93\,\Rsun\ corresponding to temperatures of 1,546\,K and 216\,K respectively \cite{brinkworthetal07-1}. This model underestimates the observed infrared flux.

As the flat, optically thick dust disc model is unable to simultaneously reproduce the presence of the nested Keplerian profiles and the observed infrared excess, we explore a more physically motivated scenario by modelling the infrared excess with the radiative transfer code MCFOST \cite{pinteetal06,pinteetal09}. Rather than a flat disc, we implemented a finite vertical structure, motivated by the fact that the dust, formed though a collisional cascade, is expected to posses a scale height comparable, or higher than the white dwarf radius \cite{reachetal09-1,Kenyon_2017-1,Balleringetal_2022}. We therefore adopted a scale height at the inner edge of the disc equal to the radius of the white dwarf ($0.015\,\Rsun$). 
In addition, the presence of gas in the disc may provide additional vertical support though gas pressure and dynamical coupling with the dust. We therefore also adopted a moderate flaring exponent of $\beta=1.2$. The dust surface density exponent is set to $p=1$ (i.e, the surface density increases outwards) to model the fact that dust is primarily produced near or beyond the Roche radius though collisional processing, while the inner regions become depleted by sublimation and collisional vaporization processes\cite{malamudetal21-1}. 
Assuming the dust to be approximately co-located with the gaseous component, we further assume a more refractory dust composition than the standard astronomical silicate mixture\cite{Draine_2003}. As a proxy for such material, we adopted amorphous carbonaceous dust optical constants\cite{RouleauMartin_1991}. Opacities were calculated using Mie theory for each grain size $a$, assuming a grain-size distribution of the form\cite{Mathis_1977} $n(a) \propto a^{-3.5}$ with grain sizes ranging from\cite{Kenyon_2017-1,Balleringetal_2022} $a_{\rm min}=0.5,\mu$m to $a_{\rm max}=5,\mu$m. 
For a disc inclination of 70\,degrees, total dust mass of $5\times10^{18}\,$g, and inner and outer edges at $0.25\,\Rsun$ and $2.0\,\Rsun$ respectively, the model is able to reproduce an IR excess comparable to the observed one, reaching a maximum temperature of $\simeq 2000\,$K at the inner edge, consistent with the sublimation temperature expected for amorphous carbon. Unlike the flat disc model, the vertically extended geometry intercepts a larger fraction of the stellar radiation through its inner rim, leading to more efficient heating of the dust. Similarly, the moderate flaring adopted in the model further increases the amount of intercepted radiation at larger radii. At an inclination of $70$\,degrees, the inner rim of the disc remains partially visible to the observer, providing an additional contribution to the near-infrared excess.
While this demonstrates that such a disc geometry can account for the observed infrared emission, it does not imply that this solution is unique. Given the large number of unconstrained parameters, the problem remains highly degenerate, and alternative combinations of disc composition, scale height, radial extent, and grain-size distribution may provide similarly acceptable fits, and potentially improve the agreement with the observed infrared excess between $1\,\mu$m and $2.5,\mu$m. 

As a final sanity check, we also modelled the IR-excess as a single blackbody located at the same distance as the white dwarf. The free parameters in this case are the temperature of the blackbody, $\mathrm{T}_\mathrm{BB}$, and its area, $\mathrm{A}_\mathrm{BB}$. We find that the SED can be reproduced by a blackbody with $T_\mathrm{BB}=1,314\pm34$\,K and an area of $2.81 \pm0.13 \times 10^{21} \mathrm{cm}^{2}$. This area corresponds to $\simeq 20$\,\% the area of the Sun or $\simeq 20$ times that of Jupiter. The presence of such a large, but undetected body is clearly unrealistic confirming that the thermal infrared flux mainly originates from dust. Though unrealistic, this  model can also be used to calculate the fractional infrared brightness of the debris disc ($\tau$) as the ratio of the integrated blackbody flux and the white dwarf flux, and compare it to the published values for other similar systems. For WD\,J1328$-$1450 we find $\tau = 2.7$\,\%, one of the largest values among all know white dwarfs with dusty debris \cite{gentilefusilloetal21-1d, dennihyetal20-2d, rocchettoetal15}.

\clearpage

\section*{Extended data}

\begin{extendtable}[h!]
\centering
\caption{\label{tab:atm-param} \textbf{Fundamental parameters of WD\,J1328$-$1450.} Top: Atmospheric parameters, stellar mass and white dwarf cooling age. Bottom: photospheric abundances, metal masses within the convection zone, diffusion time scales at the bottom of the convection zone and mass flow rates assuming accretion-diffusion equilibrium. Below we list the detected metal lines used to estimate abundances.} 
\begin{tabular}{lrcrc}
\hline 
\Teff\ [K]                   & \multicolumn{2}{l}{$17,424 \pm 567$} & \\
\logg\ (cgs)                 & \multicolumn{2}{l}{$7.76\pm 0.19$}   & \\
$M_\mathrm{WD}$ [\Msun]                 & \multicolumn{2}{l}{$0.47\pm 0.09$}& \\
$\tau_{\mathrm{cool}}$ [Myr] & \multicolumn{2}{l}{$73_{-26}^{+49}$} &\\
\hline 
 & & & & \\
\hline \smallskip
\textbf{Element} & {\bf $\log\mathrm{(Z/He)}$} & \textbf{$M_{\mathrm{Z}}$}& \textbf{$\tau$} & \textbf{$\dot{M}_{\mathrm{Z}}$}\\
& & ($10^{20}$\,g)&  $10^{5}$yr &  $10^{7}$\,$\mathrm{g/s}$\\
\hline
Mg & $-7.00\pm0.30$ & 2.94 & 4.64 & 2.01 \\
Si & $-6.30\pm0.15$ &17.04& 4.22 & 12.79\\
Ca & $-7.50\pm0.15$ & 1.53 & 3.25 & 1.50\\
\hline
& Wavelength (\AA)\\
\hline
\Ion{Mg}{II} & 4,480.7\vspace{0.2cm}\\

\Ion{Si}{II} &3,855.9\\
&3,861.9\\
&4,127.8\\
&5,041.6\\ 
&5,055.6\\ 
&6,346.2\\ 
&6,371.1\vspace{0.2cm} \\
\Ion{Ca}{II} & 3,735.9\\
& 3,932.9\\
& 3,967.9\\
\hline
\end{tabular}
\end{extendtable}

\begin{extendtable}[t]
\centering
\caption{\label{tab:emission} \textbf{Resolved emission lines in WD\,J1328$-$1450.} List of all emission lines that could be clearly resolved from the continum in the spectrum of WD\,J1328$-$1450} 
\begin{tabular}{lcc}
\hline
\textbf{Element} & \textbf{Wavelength (\AA)} & \textbf{Equivalent Width (\AA)}\\
\hline
\Ion{Fe}{II} & 5169 & 2.94$\pm$0.06\\
\Ion{O}{I} & 7774 & 7.57$\pm$0.13\\
\Ion{O}{I} & 8445 & 7.91$\pm$ 0.14\\
\Ion{Ca}{II} &  8497 & 12.90$\pm$0.13\\
\Ion{Ca}{II} & 8542 & 19.49$\pm$ 0.13 \\
\Ion{Ca}{II} & 8661 & 16.90$\pm$0.12\\
\Ion{Mg}{I} & 8806 & 4.66$\pm$0.16\\

\hline
\end{tabular}
\end{extendtable}

\setcounter{table}{2}
\begin{sidewaystable}
\centering
\caption{\textbf{Model parameters used to fit the emission line profiles.} The inner boundary of the prior distributions for $\mathrm{R_{in,1}}$ is set at the white dwarf radius, while the outer boundary for both discs is based on the locations of the emission line peaks assuming an edge-on configuration (i.e. $i=90^{\circ}$). The errors shown in column three correspond to the $16^\mathrm{th}$ and $84^\mathrm{th}$ percentiles of the MCMC posterior distributions. Values marked with (!) denote parameters that are poorly constrained given the current data.} 
\label{tab:Ca2_triplet}
\begin{tabular}{lcccccccc}
\hline
&\multicolumn{2}{c}{\Ion{Ca}{ii}} & \multicolumn{2}{c}{\Ion{O}{i}}& \multicolumn{2}{c}{\Ion{Mg}{i}}\\
\hline

Parameter&Prior (U)& Best fit&  Prior (U) & Best fit & Prior (U)& Best fit\\
\hline& \\[-1.5ex]
$\mathrm{R_{in, A}}\,[\mathrm{R_{\odot}}]$  &$(0.014,1)$&$0.14^{+0.01}_{-0.02}$ & $(0.014,1)$& $0.13^{+0.02}_{-0.04}$ & $(0.014,1)$ &$0.06^{+0.04}_{-0.03}$\\[4pt]
$\mathrm{R_{out, A}}\,[\mathrm{R_{\odot}}]$ &$(\mathrm{R_{in, A}},1)$&$0.34^{+0.02}_{-0.05}$ &$(\mathrm{R_{in, A}},1)$& $0.34^{+0.11}_{-0.04}$ &$(\mathrm{R_{in, A}},1)$&$0.31^{+0.01}_{-0.01}$\\[4pt]
$\mathrm{R_{in, B}}\,[\mathrm{R_{\odot}}]$  &$(\mathrm{R_{out, A}},5)$&$0.54^{+0.44}_{-0.16}$ &$(\mathrm{R_{out, A}},12)$& $1.21^{+0.08}_{-0.14}$&--&--\\[4pt]
$\mathrm{R_{out, B}}\,[\mathrm{R_{\odot}}]$ &$(\mathrm{R_{in, B}},5)$&$1.93^{+0.1}_{-0.3}$ & $(\mathrm{R_{in, B}},12)$&$5.14^{+4.62}_{-3.42}\,(!)$ &--&--\\[4pt]
$\alpha_A$ & $(-5,5)$&$2.39^{+0.19}_{-0.19}$ & $(-15,15)$&$2.39^{+2.06}_{-1.57}$ &$(-5,5)$&$1.79^{+0.43}_{-0.40}$\\[4pt]
$\alpha_B$ & $(-5,5)$&$-1.72^{+0.81}_{-0.56}$ & $(-15,15)$&$9.19^{+3.57}_{-7.46}$\,(!)&--&--\\[4pt]
$\zeta_{8662}$ & $(0,5)$&$0.76^{+0.01}_{-0.01}$ &--&--&--&--\\[4pt]
$\zeta_{8542}$ & $(0,5)$&$0.68^{+0.01}_{-0.01}$ &--&--&--&--\\[4pt]
$\zeta_{8498}$ & $(0,5)$&$0.89^{+0.02}_{-0.02}$ &--&--&--&--\\[4pt]
$\zeta_{O}$  & --& --&$(0,5)$&$0.73^{+0.03}_{-0.03}$ &-- &-- \\[4pt]
K$_{blend}$    & $(0,5)$&Comparison of Ca$0.61^{+0.02}_{-0.02}$ & -- &--& -- &--\\
K$_{12}$ & --& --& $(0,5)$& $0.83^{+0.29}_{-0.23}$ &-- &-- \\[4pt]
K$_{13}$ & --& --& $(0,5)$& $0.49^{+0.18}_{-0.17}$ & -- &--\\
\\[-1.5ex]
\hline& \\[-1.5ex]
$v_{sys}$ [Km/s] & $(-50,50)$&$-9.39^{+1.03}_{-0.98}$ &  \\[4pt]
$i$ [degrees]    & $\cos(i)\sim (0,1)$&$77.01^{+9.07}_{-13.47}$ & \\
\\[-1.5ex]
\hline
&  legend & &\\
\hline
$\mathrm{R_{in, A}}$ & \multicolumn{3}{l}{inner radius of the inner disc}\\
$\mathrm{R_{out, A}}$ & \multicolumn{3}{l}{outer radius of the inner disc}\\
$\mathrm{R_{in, B}}$ & \multicolumn{3}{l}{inner radius of the outer disc}\\
$\mathrm{R_{out, B}}$ & \multicolumn{3}{l}{outer radius of the the outer disc}\\
$\alpha$  & \multicolumn{3}{l}{exponent of the disc brightness profile}\\
$\zeta_{8662}$ & \multicolumn{3}{l}{Flux ratio $\zeta$ for \Ion{Ca}{ii} line at 8662 \r{A}} \\[4pt]
$\zeta_{8542}$ & \multicolumn{3}{l}{Flux ratio $\zeta$ for \Ion{Ca}{ii} line at 8542 \r{A}} \\[4pt]
$\zeta_{8498}$ & \multicolumn{3}{l}{Flux ratio $\zeta$ for \Ion{Ca}{ii} line at 8498 \r{A}}\\[4pt]
$\zeta_{O}$& \multicolumn{3}{l}{Flux ratio $\zeta$ for \Ion{O}{i} lines\r{A}}\\[4pt]
K$_{blend}$ & \multicolumn{3}{l}{Flux ratio between \Ion{Ca}{ii}\,$8498$\,\r{A} and \Ion{Ca}{ii}\,$8542$\,\r{A} lines}\\[4pt]
K$_{12}$ & \multicolumn{3}{l}{Flux ratio between \Ion{O}{i}\,$7774$\,\r{A} and \Ion{O}{i}\,$7771$\,\r{A} lines}\\[4pt]
    K$_{13}$ & \multicolumn{3}{l}{Flux ratio between \Ion{O}{i}\,$7775$\,\r{A} and \Ion{O}{i}\,$7771$\,\r{A} lines}\\
\hline
\end{tabular}
\end{sidewaystable}

\clearpage 
\begin{extendfigure}
\centering
\includegraphics[width=\hsize]{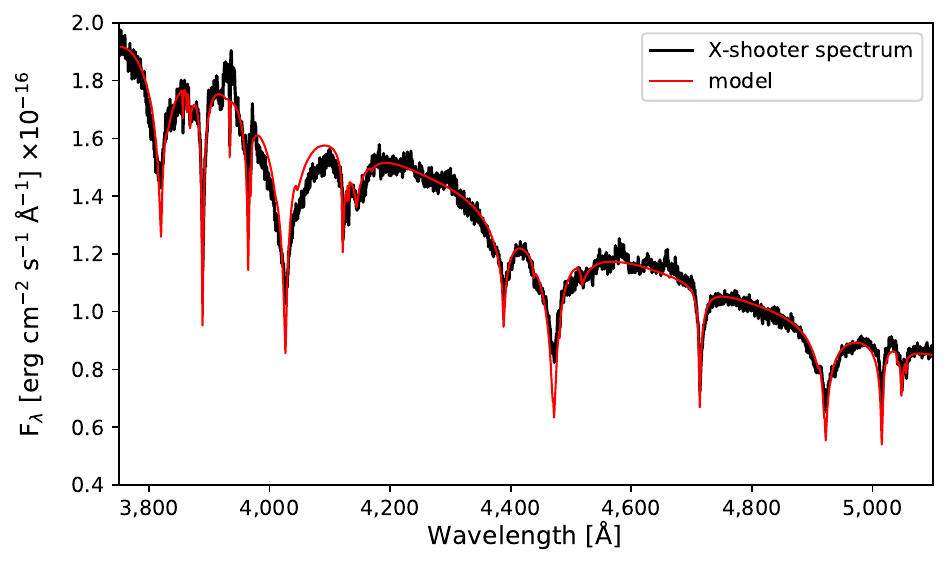}
\caption{X-shooter spectrum and best fitting model of WD\,J1328$-$1450. Best fitting model used to estimate stellar parameters and metal abundances shown in overlay on the X-shooter spectrum.}
\label{fig:model_fit}
\end{extendfigure}

\begin{extendfigure}
\centering
\includegraphics[width=\hsize]{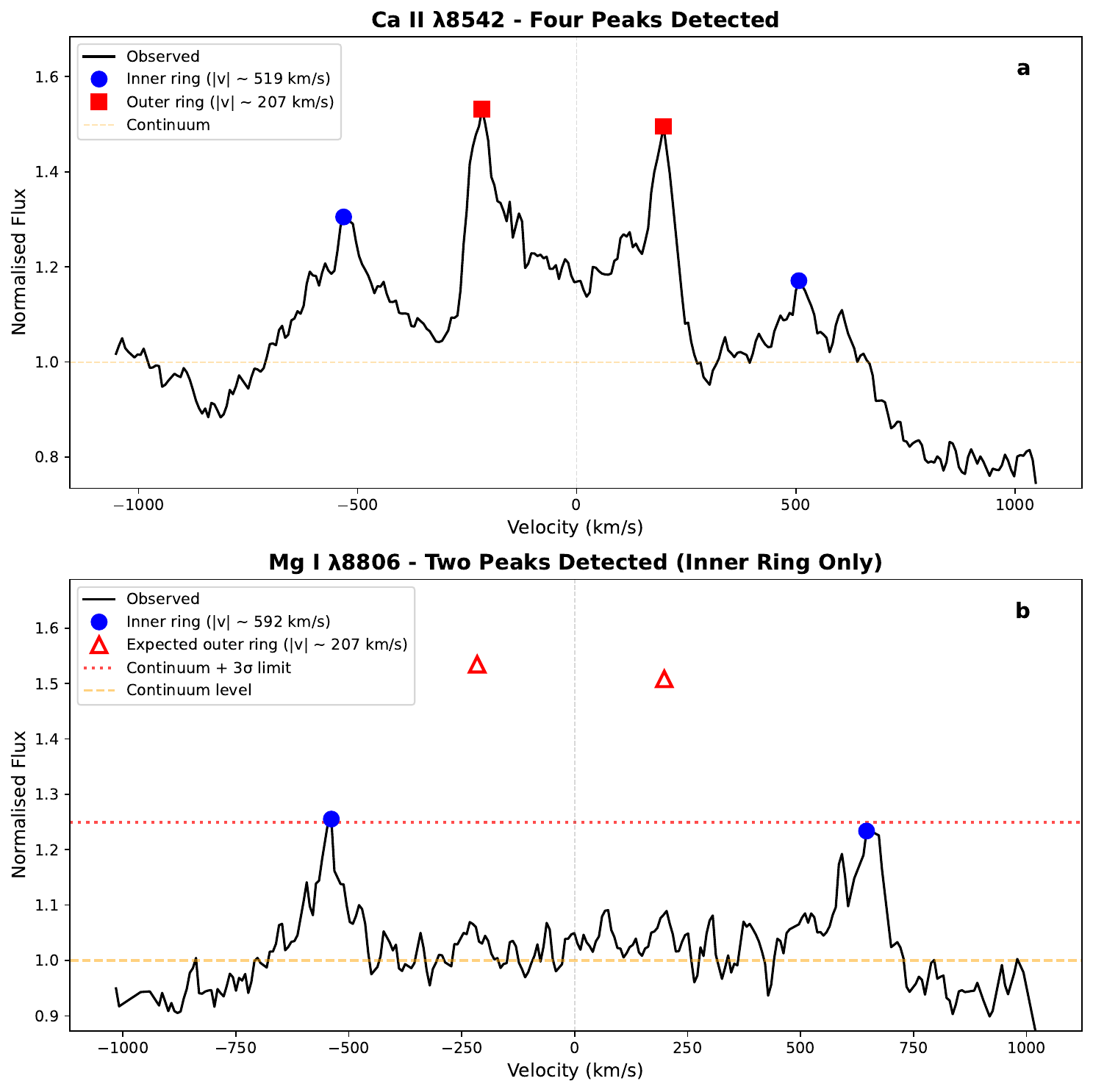}
\caption{Comparison of \Ion{Ca}{ii} and \Ion{Mg}{i} emission line profiles from the inner and outer discs. (a) The \Ion{Ca}{ii} 8542 \AA\- line shows four distinct peaks, with two peaks originating from the inner disc and two from the outer disc. (b) The \Ion{Mg}{i} 8806 \AA\- line shows only two peaks originating from the inner disc. Red triangles mark the predicted peak locations for \Ion{Mg}{i} emission from the outer disc, calculated assuming the outer disc has the same Ca/Mg abundance ratio as the inner disc. The absence of \Ion{Mg}{i} emission at these predicted locations indicates significant magnesium depletion in the outer disc.}
\label{fig:compare_ca_mg}
\end{extendfigure}

\section*{Data availability}
The X-shooter spectra analysed in the paper (see Methods sections "X-shooter observations and data reduction", "White dwarfs Parameters", "Emission lines and double disc model", as well as Extended Data Figure\,2.) are publicly available via the ESO archive \url{https://archive.eso.org/eso/eso_archive_main.html}.

\section*{Code availability}
The Koester model atmosphere code \cite{koester:2010} is not available as it requires substantial expert training and the authors do not have the resources to support general public use.
The code used to determine the discs parameters 
was made specifically for this particular system and its use requires iterative manual corrections. Therefore the code is not intended as a general-purpose tool and is made available from the corresponding author on reasonable request.

\section*{Acknowledgements}
We thank the anonymous referees for their constructive review.
Based on observations collected at the European Southern Observatory under ESO programmes 105.20G8.001, 107.22RX.001, 111.24T9.001.
This research made extensive use of Astropy (http://www.astropy.org) a community-developed core Python package for Astronomy \cite{astropy:2013, astropy:2018}. 
\section*{Funding statement}
This work was supported by the "Rita Levi Montalcini" Programme of the Ministry of University and Research (MUR).
This project has received funding from the European Research Council (ERC) under the European Union’s Horizon 2020 research and innovation programme (Grant agreement No. 101020057).
FLV acknowledges support from ANID FONDECYT Postdoctoral Grant No. 3250464
\section*{Author contributions} 
N.P. G.F.initially identified the target, carried out the spectroscopic observations and the data reduction, contributed to the model fitting of the spectrum, to the modelling of the IR excess, and lead the scientific interpretation. F. L.-V. lead the modelling of the quadruple peaked emission lines with a double disk model, contributed to the modelling of the IR excess and contributed the scientific interpretation and manuscript drafting. B. T. G. and C. J. M. contributed to the scientific interpretation and to the manuscript drafting. D. K. lead the atmospheric modelling of the star. 

\section*{Competing interests statement} 
The authors declare no competing interests.


\clearpage

\section*{Supplementary information}
\subsection*{Alternative scenarios causing the double disc structure}

\subsubsection*{A double white dwarf system.} 
The relatively low  white dwarf mass estimate $M_\mathrm{WD}=0.47 \pm 0.09$\,\Msun raises the possibility that  WD\,J1328$-$1450 could be a wide, but unresolved double degenerate system hosting a debris disc around each component and exhibiting therefore the quadruple peak configuration in the combined spectrum. 
WD\,J1328$-$1450 lies at a distance of $\simeq500$\,pc and Gaia reports only a single source. The angular resolution limit of Gaia for reliably resolving binary companions is approximately\cite{Lindegren_2018} 0.3\,arcsec, corresponding to a projected physical separation of $\simeq150$\,au at the system's distance. 
If WD\,J1328$-$1450 were an unresolved double white dwarf binary with current separation $\lesssim 150$\,au, the  main-sequence progenitor binary would have been significantly more compact. During the asymptotic giant branch phase, the stars would have lost $50-80$\% of their mass causing the binary orbit to expand by a factor of $\simeq2-5$ \cite{Verasetal11}. Working backwards, the pre-mass-loss separation would have been $\simeq30 - 75$\, au.
For planets orbiting individual components of a binary (circumstellar orbits), stability demands $a_\mathrm{planet} \lesssim 0.3 - 0.5 \times a_\mathrm{binary}$ , where $a$ indicates the semimajor axis of the orbit \cite{Holman_Wiegert99}
Hence, planets orbiting each white dwarf individually would need their orbits to be within $10 - 35$\,au to remain stable. Even though these orbits are plausible, and these planets could survive the stars evolution into white dwarfs, the double disc configuration would require additional coincidental occurrences. Only 0.067\% of white dwarfs show gaseous debris discs \cite{manseretal20-1} so the odds of two occurring independently, but simultaneously for the two components of a binary system are extremely low. This, combined with the rarity of close white dwarf binaries hosting dual planetary systems \cite{Farihi16, Zuckerman14}, makes this scenario unlikely. Moreover, delivering material from surviving planets at these distances to the observed sub-solar-radius disc locations requires fine-tuned dynamical architectures \cite{Bonsor_Veras15, Verasetal24}, compounding the improbability. We therefore favour the interpretation of single white dwarf hosting two discs.

\subsubsection*{A single planetesimal in the disc gap.} 
We considered an alternative scenario where the double disc structure observed in WD\,J1328$-$1450 may be the result of single planetary body creating a gap both in the dust and the gas disc.  This interpretation is broadly analogous to what seen in protoplanetary discs where a sufficiently large planetary body can clear a path in a dust disc by accreting the material along its orbit. Though initially this picture may appear as the simplest solution, it faces a number of challenges.  As mentioned in the main text, assuming that the gap is also present in the dust disc exacerbates the issue of the IR-excess being too large for a simple geometrically-thin optically-thick disc.  Furthermore, while one could imagine a large body clearing the dust by directly colliding into it, removing the gas would be far more difficult. The planet would need enough mass to accrete and trap the gas. At 0.62\,\Rsun, even assuming the planet is entirely made of iron ($\rho = 7.9\,\mathrm{g\,cm^{-3}}$) and its internal strength is at the upper bound of what observed for iron bodies (500\,Mpa), its maximum size before undergoing tidal disruption would be 440\,km. Even in this extreme scenario the escape velocity from the surface of this body would be only $\simeq 1.4$\,m/s, orders of magnitude below the thermal velocity of the gas in the disc assuming a conservative temperature of $\simeq 1,000$ K.  Finally, in this scenario the planet in the gap would not directly interact with the dust in the two discs on either side of the gap so a separate mechanism would be required to generate the gas in the two discs.

\subsubsection*{Disc self-shadowing.}
The quadruple emission features seen in WD\,J1328$-$1450 could, in theory, also be explained without invoking a physical gap in the gas distribution. If the inner radius of a flaring disc is truncated at the sublimation radius then its inner rim, being directly irradiate by the white dwarf, will acquire a higher temperature and vertical scale height compared to a fully-flaring or flat geometry. This puffed inner region will cast a shadow \cite{Dullemond_2001} over part of the disc surface behind it preventing the photoionization of the gas. This scenario, applied to WD\,J1328$-$1450, would suggest that the puffed inner region is responsible for the line emission between $0.14-0.34\,\Rsun$, the shadowed region would be located between $0.34-0.54\,\Rsun$ and the outer part of the disc beyond $0.54\,\Rsun$ would have a sufficient vertical height to be out of the shadow and once again exposed to irradiation.

In addition to explaining the observed quadruple emission feature, the vertical extent of the disc and the lack of a need for a gap of material between $0.34-0.54\,\Rsun$ imply a larger amount of dust which would explain the observed high infrared excess. This scenario also naturally explain the opposite behaviour of the brightness profiles derived for both discs. The amount of ionized gas in the puffed region, being directly irradiate by the white dwarf, decreases according to the inverse-square law ($\alpha_1\gtrsim2$), while the gas in the outermost non-shadowed region of the disc would progressively intercept more radiation as we move further away from the white dwarf as a consequence of the flared geometry (i.e. $\alpha_2 \lesssim 0$).

However, this interpretation is not without its flaws. Firstly, one needs to question why no other white dwarf with gaseous debris, and even more no other astronomical objects with circumstellar gas discs and double peaked emission lines (like Herbig Ae/Be and T Tauri stars \cite{Thi_2001, Acke_2005, Nidhi_2023}), has ever been observed to have quadruple peaked emission lines.
This implies that, for self-shadowing to be plausible, WD\,J1328$-$1450 must posses some intrinsic unique properties. For example, the dusty debris disc around WD\,J1328$-$1450 is very bright, suggesting a geometry more complex than a simple flat disc which could be a contributing factor to self-shadowing. However, a number of systems with even brighter discs exist. For example WD\,J0846+5703 has a fractional infrared brightness over twice as large as that of WD\,J1328$-$1450, but only shows the typical double peaked emission lines \cite{gentilefusilloetal21-1d}. Alternatively, the flux from the white dwarf and/or the chemical composition of the disc may be the most significant factors in enabling the self shadowing scenario, but again WD\,J1328$-$1450 is not special in these regards. In terms of \Teff\,  and overall luminosity WD\,J1328$-$1450 sits in the middle of the distribution of known gas disc hosts, so the  level of irradiation of the disc should not be in any way atypical. Similarly WD\,J1328$-$1450, like many other white dwarf with gas discs, displays emission lines of Fe, O, Mg, and Ca and so does not appear to host debris with a peculiar chemical composition. In contrast the gas disc host WD\,J0529$-$3401  exhibits over 50 emission lines from a wider range of elements indicative of a more peculiar debris chemical composition, but all these lines only display the typical double peaked profile \cite{gentilefusilloetal21-1d}.  

In conclusion, though disc self shadowing can theoretically explain the quadruple peak line profiles in WD\,J1328$-$1450, no direct evidence of this phenomenon exists and there are no observations or arguments to explain why this mechanism would be present only here and not in other similar systems.

\begin{sidewaystable}
\centering
\caption{\textbf{Astrometry and Photometry of WD\,J1328$-$1450.} The optical photometry above the dashed line was used in the fit of the white dwarf parameters (see Methods, White Dwarf Stellar Parameters), the data below that line was only included in the analysis of the infrared excess (see Methods, infrared excess and dust disc).} 
\label{tab:photo}
\begin{tabular}{llcccc}
\hline
\textbf{Survey}& \textbf{Parallax [mas]} & \textbf{Uncertainty [mas]}\\
\hline
\textit{Gaia}  & 2.014  & 0.292\\
\noalign{\bigskip}
\hline
\hline

\textbf{Survey}& \textbf{Filter}& \textbf{Effective Wavelength [\AA] } & \textbf{Magnitude} & \textbf{Uncertainty}\\
\hline
SkyMapper & $u$ & 3,500.22 & 18.792 & 0.058 \\
~~~~"     & $v$ & 3,878.68 & 18.776 & 0.013 \\
~~~~"     & $g$ & 5,016.05 & 18.800 & 0.028 \\
~~~~"     & $r$ & 6,076.85 & 19.040 & 0.021 \\
~~~~"     & $i$ & 7,732.83 & 19.336 & 0.013 \\
~~~~"     & $z$ & 9,120.25 & 19.369 & 0.062 \\[0.5ex]
PAN-Starrs (PS1) & $g$ & 4,810.16 & 18.767 & 0.014\\
~~~~"            & $r$ & 6,155.46 & 19.043 & 0.006\\
~~~~"            & $i$ & 7,503.03 & 19.339 & 0.012\\
~~~~"            & $y$ & 9,613.60 & 19.549 & 0.055\\[0.5ex]
\hdashline
VHS      & $J$           & 12,480.62 & 19.062 & 0.126 \\
~~~~"    & $H$           & 16,347.44 & 18.511 & 0.116 \\
~~~~"    & $K\mathrm{s}$ & 21,435.27 & 17.737 & 0.137 \\
catWISE  & $W1$          & 33,154.27   & 16.577 & 0.043 \\
~~~~"    & $W2$          & 45,644.77 & 15.911 & 0.067 \\
\hline
\end{tabular}
\end{sidewaystable} 
\clearpage

\begin{sidewaystable}

\centering
\caption{\label{tab:line_properties}\textbf{Air wavelengths, fluxes, equivalent widths, and peak-to-peak separations of the inner and outer disc profiles for each line fitted with the double disc model}. The peak-to-peak separations for the inner ($\Delta V_\mathrm{in,peak}$) and outer ($\Delta V_\mathrm{out,peak}$) disc profiles are estimated from the best MCMC models, with errors accounting for uncertainties in the white dwarf mass, outer disc radius, and inclination. Since the \Ion{Ca}{ii} triplet was fitted with a single double disc model, the measured peak-to-peak separations are the same for the three lines.} 
\begin{tabular}{llcccc}
\hline

Element&Wavelength [\AA]& Flux [$10^{-16}\mathrm{erg\,cm^{-2}\,s^{-1}}$]& EW [\AA] & $\mathrm{\Delta V_{in,peak}}$ [$\mathrm{km\,s^{-1}}$] & $\Delta V_\mathrm{out,peak}$ [$\mathrm{km\,s^{-1}}$] \\
\hline
\Ion{Ca}{ii} & 8,498 & $2.10\pm 0.04$ & $12.9 \pm 2.2$ & $1,024 \pm 120$ & $420 \pm 52$ \\[0.5ex]
\Ion{Ca}{ii} & 8,542 & $3.10\pm 0.04$ & $19.5 \pm 2.0$ & " & "\\[4pt]
\Ion{Ca}{ii} & 8,662 & $2.64\pm 0.04$ & $16.9 \pm 3.0$ & " & " \\[4pt]
\Ion{Mg}{i}  & 8,806 & $0.77\pm 0.04$ & $5.3 \pm 3.9$ & $1,072 \pm 51$ & - \\[4pt]
\Ion{O}{i}   & 7,774 & $1.90\pm 0.05$ & $9.2 \pm 2.2$ & $1,036 \pm 188$ & $470 \pm 120 $ \\
\hline
\end{tabular}
\end{sidewaystable}

\clearpage 

\begin{suppfigure}
\centering
\includegraphics[width=\hsize]{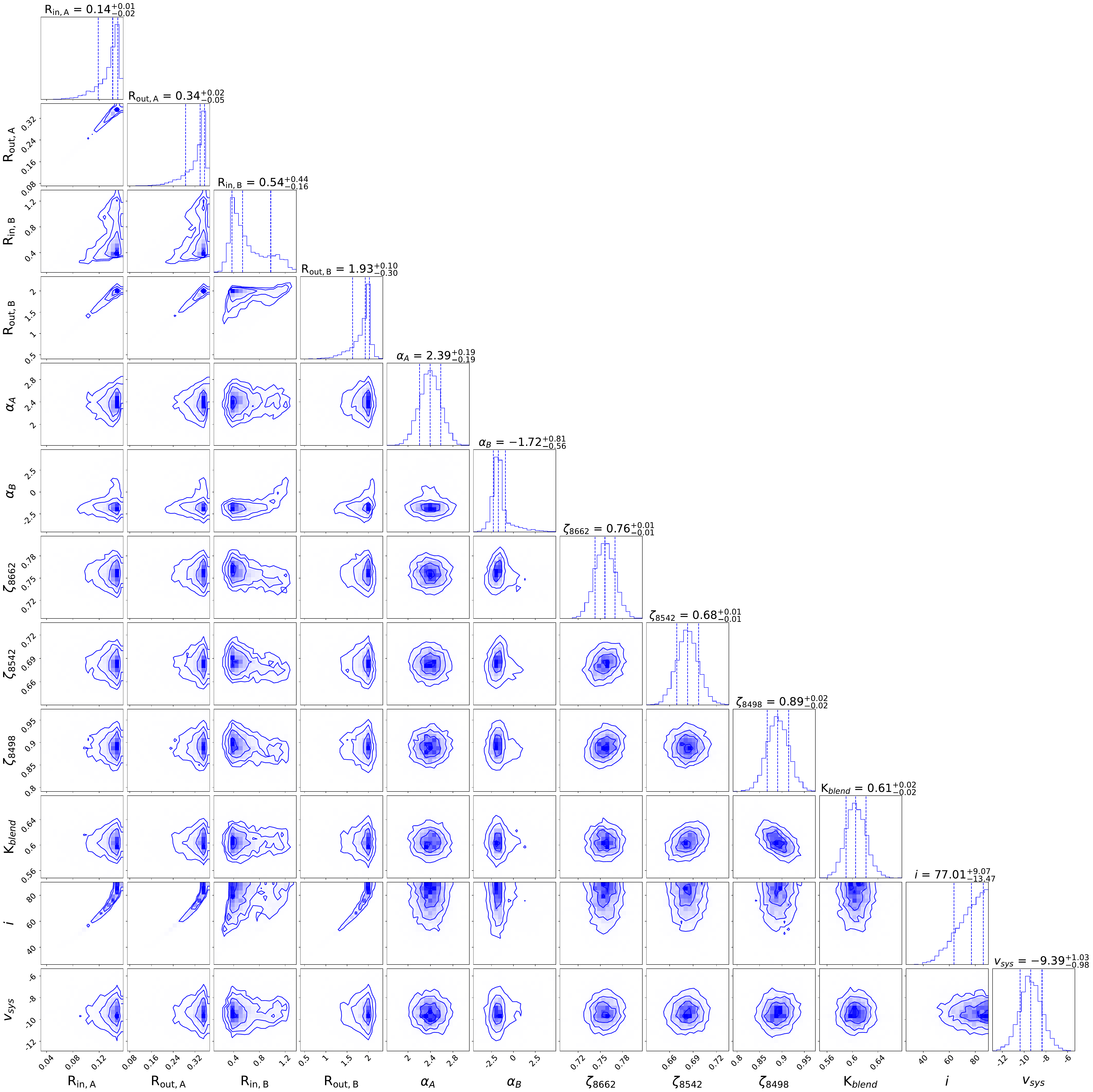}
\caption{\textbf{Corner plot diagram for the fit of the \Ion{Ca}{ii} triplet.} The diagram shows the covariance and posterior distribution of the free parameters of our double-disc model. Vertical-dashed lines marks from left to right the $16^\mathrm{th}$, $50^\mathrm{th}$ and  $84^\mathrm{th}$ percentiles.}
\label{fig:_corner_O1_triplet}    
\end{suppfigure}

\begin{suppfigure}
\centering
\includegraphics[width=\hsize]{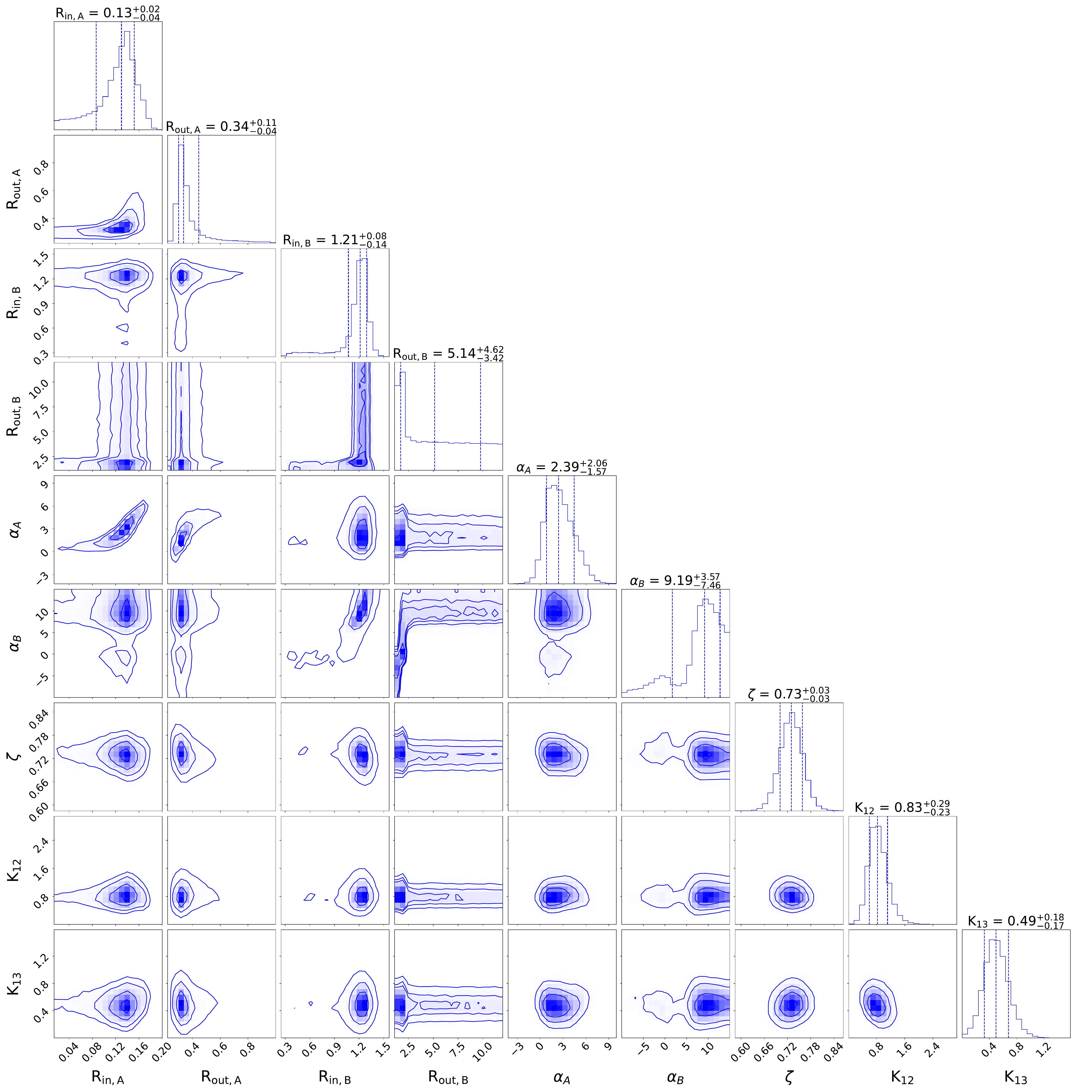}
\caption{\textbf{Corner plot diagram for the fit of the \Ion{O}{i} triplet.} The diagram shows the covariance and posterior distribution of the free parameters of our double-disc model. Vertical-dashed lines marks from left to right the $16^\mathrm{th}$, $50^\mathrm{th}$ and  $84^\mathrm{th}$ percentiles. Even though most of the parameters are relatively well constrained, the blending of the lines conspires against a more accurate characterization of the nested profiles. This is evident for $\mathrm{R_{out,2}}$ and $\alpha_2$ as the emission lines produced by the outer disc are located at the center of the triplet.}
\end{suppfigure}

\begin{suppfigure}
\centering
\includegraphics[width=0.5\hsize]{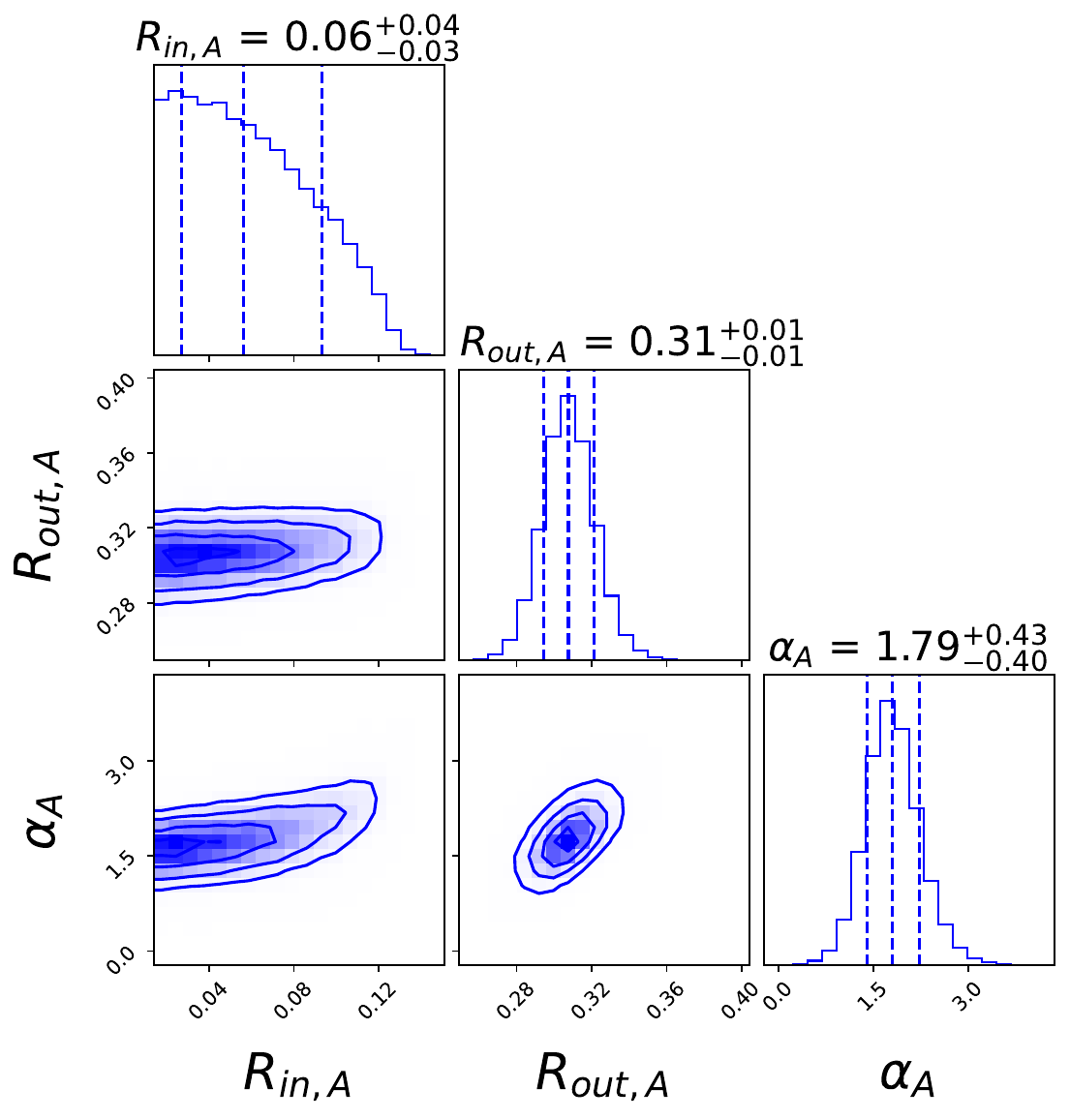}
\caption{\label{fig:_corner_O1_triplet2}\textbf{Corner plot diagram for the fit of the of the \Ion{Mg}{i} emission line}.  The diagram shows distribution of the parameters in the fit of the double-peaked profile.}
\end{suppfigure}

\end{document}